\documentclass{article}

\usepackage{arxiv}

\usepackage[utf8]{inputenc} 
\usepackage[T1]{fontenc}    
\usepackage[hidelinks]{hyperref}
\usepackage{url}            
\usepackage{booktabs}       
\usepackage{amsfonts}       
\usepackage{nicefrac}       
\usepackage{microtype}      
\usepackage{graphicx}
\usepackage{natbib}
\usepackage{doi}
\usepackage{amsmath}
\usepackage{placeins}
\usepackage{xcolor}
\usepackage{authblk}

\title{Magnetic Reconnection and High-Speed Jets Downstream of a Parallel Shock: Full Particle-in-cell Simulations}
\date{}
\author[1,2]{Wanxin Ouyang}
\author[1]{Zhongwei Yang \textsuperscript{*}}
\author[3]{Quanming Lu \textsuperscript{*}}
\author[4]{Gary P. Zank}
\author[5]{Heli Hietala}
\author[6]{Shuichi Matsukiyo}
\author[7]{Can Huang}
\author[1]{Binbin Tang}
\author[1]{Wence Jiang}
\author[1]{Tianran Sun}
\author[1,2]{Xiaocheng Guo}
\author[1,2]{Hui Li}
\author[1,2]{Chi Wang}
\affil[1]{State Key Laboratory of Solar Activity and Space Weather, National Space Science Center, CAS, Beijing, 100190, China (zwyang@swl.ac.cn)}
\affil[2]{University of Chinese Academy of Sciences, Beijing, 101408, China}
\affil[3]{CAS Key Lab of Geospace Environment, School of Earth and Space Sciences, University of Science and Technology of China, Hefei, 230026, China (qmlu@ustc.edu.cn)}
\affil[4]{Center for Space Plasma and Aeronomic Research (CSPAR), University of Alabama in Huntsville, Huntsville, AL 35805, USA}
\affil[5]{Department of Physics and Astronomy, Queen Mary University of London, London, London E1 4NS, UK}
\affil[6]{Faculty of Engineering Sciences, Kyushu University, Kasuga, Fukuoka 816-8580, Japan}
\affil[7]{CAS Engineering Laboratory for Deep Resources Equipment and Technology, Institute of Geology and Geophysics, CAS, Beijing, 100029, China}

\renewcommand{\shorttitle}{Magnetic Reconnection and High-Speed Jets}
\hypersetup{
  pdftitle={Magnetic Reconnection and High-Speed Jets Downstream of a Parallel Shock: Full Particle-in-cell Simulations},
  pdfauthor={Wanxin Ouyang, Zhongwei Yang, Quanming Lu, Gary P. Zank, Heli Hietala, Shuichi Matsukiyo, Can Huang, Binbin Tang, Wence Jiang, Tianran Sun, Xiaocheng Guo, Hui Li, Chi Wang},
  pdfsubject={Full particle-in-cell simulations of a parallel collisionless shock},
  pdfkeywords={Shock, Particle acceleration, High-speed jets, Magnetic reconnection}
}

\begin{document}
\raggedbottom
\maketitle

\renewcommand{\thefootnote}{}
\footnotetext{Corresponding authors: Zhongwei Yang$^{*}$ (\href{mailto:zwyang@swl.ac.cn}{zwyang@swl.ac.cn}), Quanming Lu$^{*}$ (\href{mailto:qmlu@ustc.edu.cn}{qmlu@ustc.edu.cn}).}
\renewcommand{\thefootnote}{\arabic{footnote}}

\begin{abstract}
Collisionless shocks are of fundamental interest in astrophysics and space physics as important sources of energetic particles. Downstream of quasi-parallel shocks, high-speed jets (HSJs) and magnetic reconnection have been frequently observed, while it remains unclear how their coupling shapes structured regions for particle energization. We use two-dimensional full particle-in-cell simulations to investigate HSJ generation, thin current sheet formation, and associated electron energization downstream of a parallel shock. We find that portions of foreshock ULF waves with enhanced dynamic pressure interact with a self-reforming shock front and are transmitted downstream, forming HSJs, while neighboring low pressure portions evolve into cavities. At the leading edge of the HSJs, the unique bent magnetic topology carried by the compressed ULF waves produces bow waves and locally squeezes out thin current sheets. These current sheets show reconnection signatures, including X-point topology, magnetic islands, enhanced $J\cdot E'$, and Alfvénic outflows along the local reconnecting direction. In the reconnection region associated with HSJs, the parallel electric field, betatron, and Fermi terms all contribute to adiabatic electron energization. The betatron term appears a visible signature near the X line, where the local magnetic structure is compressed by the HSJ. For the first time, we report a systematic mass-ratio dependence of HSJ morphology in full PIC simulations: lower mass ratios produce fragmented HSJs with shorter lifetimes, whereas larger mass ratios yield longer and wider HSJs with clearer bow waves as ion inertia increases.
\end{abstract}


\section{Introduction} \label{sec:s1}

Collisionless shocks and their surrounding plasma environments contain rich wave activity, extending from the upstream all the way to the downstream \citep{Lembege2004,Burgess2005}. On localized scales, the shock front is often nonstationary, as predicted by kinetic simulations and confirmed by MMS observations, with common forms including shock front rippling \citep{Winske1988,Johlander2016,Gingell2017}, shock front self-reformation driven by reflected ions \citep{Lembege1987,Scholer2003,Su2012,Yang2020a}, and reformation associated with the interaction of upstream ULF waves \citep{Liu2020a,Turner2021,Zhang2025} or whistler waves \citep{Pantellini1992,Matsukiyo2003,Matsukiyo2024,Xu2026} with the shock surface. These nonstationary shock-front structures and their associated upstream and downstream transients can strongly modulate plasma compression, field-line distortion, and current-sheet formation near the shock.

Magnetosheath high-speed jets (HSJs) are typical transient structures commonly observed downstream of quasi-parallel shocks \citep{Hietala2009,Plaschke2018,Escoubet2020}. Early observations of such structures in the terrestrial magnetosheath were reported by \citet{Nvemevcek1998}, and subsequent studies have reported HSJs in the Martian and Jovian magnetosheaths \citep{Gunell2023,Zhou2024}, possible analogues in the Hermean magnetosheath \citep{Karlsson2016}, and HSJs downstream of interplanetary shocks \citep{Hietala2024}. HSJs are typically characterized by dynamic pressure exceeding that of the ambient magnetosheath plasma \citep{Archer2013} and approaching upstream solar wind values \citep{Plaschke2020}, together with enhanced ion density and bulk velocity \citep{Karlsson2018}. These properties distinguish HSJs from embedded plasmoids, which show density enhancements and magnetic field variations without corresponding flow enhancement. Embedded paramagnetic plasmoids may originate from magnetic flux ropes generated by downstream magnetic reconnection \citep{Preisser2020}, whereas embedded diamagnetic plasmoids may instead have upstream solar wind origins \citep{Karlsson2015}. Extensive statistical studies have characterized the propagation, geometry, and lifetime evolution of HSJs \citep{Karlsson2018,Raptis2020,Lamoury2021,Wang2026}, as summarized in the recent review by \citet{Kramer2025}. Spacecraft and ground-based observations further indicate that HSJs are geoeffective, perturbing the dayside magnetopause \citep{Archer2013,Hietala2018,Guo2024,Ma2024}, triggering ground magnetic ultra-low frequency waves \citep{Wang2022}, and producing localized ionospheric convection and auroral responses \citep{Han2017,Wang2018,Hietala2012}.

The generation of jets downstream of quasi-parallel shocks is generally believed to be related to shock front nonstationarity and may involve several mechanisms, including magnetosheath jets produced by concave ripples on the shock surface \citep{Hietala2009}, jets generated by the combined effect of foreshock wave evolution and an ongoing shock front reformation cycle \citep{Raptis2022}, and moderate-scale jets produced at the expanding edges of preexisting HSJs, often appearing as paired structures \citep{Zhou2024,Guo2026}. Local hybrid \citep{Hao2016} and global hybrid simulations have reproduced magnetosheath jets downstream of the quasi-parallel region of the terrestrial bow shock, in good agreement with observations \citep[e.g.,][]{Chen2021,Suni2021,Guo2022,Yang2024,Ren2024a,Fatemi2024}. In early two-dimensional hybrid simulations, compressional bow waves (BWs) are identified at the leading edge of some magnetosheath jets \citep{Karimabadi2014}. More recently, \citet{Ren2024b} simulate the fine structures of jet-driven BWs. Their results indicate that BWs are primarily characterized by enhanced magnetic field and ion-density compression regions, with the corresponding peaks separated by approximately one ion inertial length. Using THEMIS data, \citet{Liu2019} confirm HSJ-related BWs in Earth's magnetosheath. Subsequent observations and test-particle calculations further demonstrate that a fraction of jet-associated BWs can accelerate particles and contribute to energy dissipation in the magnetosheath \citep{Liu2020b,Liu2020c,Vuorinen2022}. Two-dimensional local hybrid simulations by \citet{Ren2023} show that HSJ generation is preferentially favored under a strictly parallel shock configuration, rather than under a general quasi-parallel geometry, especially for low plasma beta or high Alfvén Mach number. Both hybrid-PIC simulations \citep{Ren2023} and hybrid-Vlasov simulations \citep{Suni2021} show that foreshock compressive structures, which are inhomogeneous perpendicular to the background magnetic field, may interact with the shock front and produce shock ripples and downstream HSJs. Further investigating the upstream origin of these compressive structures, \citet{Ren2026} suggest that the nonlinear evolution of upstream foreshock ULF waves at a parallel shock can lead to dynamic pressure inhomogeneities along the shock surface, with high dynamic pressure regions linked to the formation of downstream HSJs.

In addition to studies of HSJ formation downstream of quasi-parallel shocks, separate kinetic studies and MMS observations have shown that magnetic reconnection can also occur within the shock transition region and in the downstream magnetosheath. For quasi-parallel shocks, two-dimensional full PIC simulations by \citet{Bessho2019,Bessho2020} show that short-wavelength modes (on the order of the ion inertial length or less) propagating obliquely to the magnetic field bend the magnetic field lines, allowing magnetic reconnection to occur where oppositely directed field lines come into contact within the shock transition region. In addition to the short-wavelength kinetic modes discussed above, long-duration 2-D and 3-D kinetic simulations by \citet{Lu2020,Lu2024} examined a regime in which upstream foreshock ULF waves with wavelengths greater than about $50\ d_{i0}$ become fully developed, similar to those found in previous hybrid simulations \citep[e.g.,][]{Scholer1990,Hao2016,Sundberg2016}. In this regime, the upstream ULF waves evolve from the linear stage to nonlinear steepening as they propagate toward the shock front, eventually interacting with the shock surface and convecting downstream. During this process, the magnetic field lines associated with the ULF waves are periodically bent and compressed at the shock front, producing patchy magnetic reconnection in the immediate downstream region. A fraction of the resulting current sheets can develop ion-coupled reconnection downstream, whereas current filaments can host electron-only reconnection and chains of small-scale magnetic islands with a string-of-pearls morphology, indicating that both reconnection regimes can coexist. Observations indicate that the wave modes represented in these kinetic simulations, including large-amplitude short-wavelength whistler waves and low-frequency large-scale ULF waves, can occur under quasi-parallel shock conditions (for examples, see \citep{Wilson2016}). Reconnection signatures have also been identified downstream of quasi-perpendicular shocks \citep{Lu2021,Guo2023} and perpendicular shocks \citep{Matsumoto2015,Bohdan2020}. Corresponding evidence for strong current sheets has been reported downstream of shocks \citep{Eriksson2016}, and magnetic reconnection signatures have been observed both in the shock transition region \citep{Wang2019,Gingell2020,Jiang2021} and in the magnetosheath \citep{Phan2018,Eriksson2018,Lu2022,Wang2024}.

Although previous numerical simulations have shown that HSJs and shock-associated magnetic reconnection can both occur downstream of quasi-parallel shocks, several recent observational and simulation studies have further identified current sheets and reconnection signatures near HSJ fronts \citep{Roberts2025}. However, the causal relationship between HSJs and magnetosheath reconnection remains unresolved. In particular, it is still unclear whether HSJs can directly trigger reconnection within the magnetosheath, where such reconnection preferentially occurs relative to the jet structure, which type of reconnection develops, and how the associated particles are accelerated. These open questions motivate further event studies and fully kinetic PIC simulations that can self-consistently resolve the coupled evolution of HSJs, current sheet formation, and reconnection onset.

In this study, we employ 2-D full PIC simulations to investigate magnetic reconnection induced by a magnetosheath jet downstream of a parallel shock. The paper is organized as follows: Section 2 introduces the simulation model and numerical setup. Section 3 presents the relationship between shock reformation and the formation of HSJs, BWs, and compressional current sheets. It also examines the reconnection process triggered by the magnetosheath jets, as well as the associated particle energization. Section 4 discusses the influence of different ion-to-electron mass ratios and summarizes the main conclusions.

\section{Simulation Model} \label{sec:s2}

This paper employs an electromagnetic full PIC simulation code \citep{Lembege1987,Yang2009} to investigate magnetic reconnection induced by HSJs downstream of a collisionless parallel shock. A so-called injection method is adopted to generate a planar shock in the downstream rest frame, as in previous simulations \citep{Matsukiyo2014,Yang2014,Yang2020b,Matsukiyo2024}. The plasma is continuously injected from the left boundary ($x=0$) of the 2D simulation domain. It propagates toward the right boundary ($x=L_x$), where $L_x$ denotes the simulation domain size in the $x$ direction. The plasma is then reflected at the right boundary, generating a shock that propagates toward the left ($-x$ direction). The corresponding Alfvén Mach number $M_A$ is about 9. The simulations are performed in the 2D $x$-$y$ plane. The background magnetic field is initialized in the $+x$ direction. In this study, a parallel-shock configuration is adopted with a shock-normal angle of $\theta_{Bn}=0^\circ$. Periodic boundary conditions are applied to both electromagnetic fields and particles in the $y$ direction. For the simulations with $m_i/m_e=100$ and 400, the number of grid cells is $n_x \times n_y =40000 \times 2000$, with 64 particles per cell for each species. The corresponding simulation domains are $L_x \times L_y =1520d_{i0} \times 200d_{i0}$ and $760d_{i0} \times 200d_{i0}$, respectively, where $d_{i0}$ is the ion inertial length defined using the upstream density. The spatial resolutions are $\Delta x=0.038d_{i0}$ and $0.019d_{i0}$ in the $x$ direction, respectively, while $\Delta y=0.1d_{i0}$ in both cases. In both simulations, $\Delta x=0.38d_{e0}\approx0.98\rho_{e0}$ in the upstream region, where $d_{e0}$ and $\rho_{e0}$ are the upstream electron inertial length and electron gyroradius, respectively. The corresponding values of $c/V_{\mathrm{A}e0}$ are 6 and 3, respectively, where $V_{\mathrm{A}e0}$ is the upstream electron Alfvén speed. These resolutions adequately resolve most of the upstream fluctuations and the downstream reconnection region examined in this study. The ion and electron plasma beta values are set to $\beta_i=\beta_e=0.3$. These shock parameters are not tailored to a special event, but are representative of conditions reported for previous HSJ and nonstationary quasi-parallel shock observations \citep{Hietala2009,Johlander2022}. They are also close to those used in previous hybrid simulations of HSJ-driven bow waves \citep{Ren2024b}, allowing a direct comparison with the present full PIC results. In addition, statistical studies have shown that higher Alfvén Mach numbers favor the occurrence of magnetosheath jets and associated bow waves \citep{Liu2020}. The simulation with $m_i/m_e=400$ is adopted as the main case in this study, while the $m_i/m_e=1200$ case with $\Delta x=0.05d_i$ is shown only for reference.

\section{Simulation Results} \label{sec:s3}

\subsection{Foreshock and Downstream Transients at a Parallel Shock} \label{sec:s3.1}

Figure 1 presents a snapshot of the upstream and downstream transient structures at the simulated parallel shock at $t=54\ \Omega_{ci0}^{-1}$. Panel (a) shows an overview of the ion dynamic pressure $P_d/P_{d0}$ across the upstream, shock front, and downstream regions, where $P_{d0}$ is the far upstream reference value. Warm colors denote enhanced dynamic pressure, whereas cool colors indicate regions of reduced dynamic pressure. The profiles above Figure 1a show the $Y$-averaged magnetic field components $B_y/B_0$ and $B_z/B_0$, together with the shifted magnetic field magnitude $B/B_0$, where $B_0$ is the initial upstream ambient magnetic field strength. The horizontal bars mark the upstream (US, red), shock front (Front, yellow), and downstream (DS, blue) regions, and the vertical dashed line denotes the shock overshoot location.

In the upstream region, large-amplitude ULF waves are visible. Such foreshock ULF waves are commonly associated with backstreaming ions reflected from quasi-parallel shocks \citep{Su2012,Hoppe1981,Gary1981,Burgess2005}. Their nonlinear evolution can produce dynamic pressure inhomogeneities along the shock surface, with high dynamic pressure regions linked to the downstream emergence of HSJs, suggesting a possible scenario for their generation \citep{Ren2026}.

Downstream of the shock, several localized structures with high dynamic pressure are observed and identified as HSJs. These HSJs appear as elongated, meandering pressure enhancements embedded in the downstream region and are separated by reduced pressure regions labeled as cavities. This spatial pattern suggests that the enhanced pressure portions of upstream ULF waves are spatially associated with downstream HSJs, whereas neighboring reduced pressure portions are associated with downstream cavity-like structures. Similar alternating jet and cavity patterns have been reported in previous global hybrid simulations of quasi-parallel shocks \citep{Ren2024a,Yang2024,Fatemi2024}. Figures 1b and 1c show enlarged views of the white box in Figure 1a, displaying $B/B_0$ and $N/N_0$, respectively. Magnetic field lines are overplotted as black curves. The HSJs are accompanied by enhanced magnetic field and density, while the cavity region is characterized by reduced density and weaker magnetic field. This reduction in both density and magnetic field is consistent with the basic properties of foreshock and magnetosheath cavities reported in previous observations \citep{Sibeck2002}. The enlarged views further show strongly bent magnetic field lines around the HSJ and cavity structures, indicating a highly structured downstream magnetic topology. In the following subsection, we focus on the HSJ near $Y=70\ d_{i0}$ to examine its life cycle, including its formation, temporal evolution, and the bow wave driven at its leading edge.

\subsection{Formation mechanisms of HSJs and Bow Waves} \label{sec:s3.2}

Figure 2 focuses on the HSJ near $Y=70\ d_{i0}$ selected from Figure 1 and shows its formation and evolution during shock self-reformation. Figures 2a-2e show the ion dynamic pressure $P_d/P_{d0}$ at $t=40.2$, 44.2, 48.2, 52.9, and $56\ \Omega_{ci0}^{-1}$, respectively. As in Figure 1, the colored bars above each panel mark the upstream, shock front, and downstream regions at the corresponding time, and the white vertical dashed lines denote the shock overshoot locations. The white curve outlined in red shows the ion number density profile $N_i/N_0$ averaged from $Y=65\ d_{i0}$ to $75\ d_{i0}$ and shifted as $2.5N_i/N_0+43$ for display. The labels I, II, and III indicate successive shock overshoots during the self-reformation process. Figures 2a-2c show one representative self-reformation period. At $t=40.2\ \Omega_{ci0}^{-1}$ (Figure 2a), the shock front associated with overshoot I is relatively steep. By $t=44.2\ \Omega_{ci0}^{-1}$ (Figure 2b), the front becomes flatter, and a high dynamic pressure portion of the upstream ULF wave crosses overshoot I and enters the immediate downstream region. At $t=48.2\ \Omega_{ci0}^{-1}$ (Figure 2c), a new overshoot, labeled II, begins to form upstream of the old overshoot I, and its density peak exceeds that of the old front. During this interval, the sinuous high dynamic pressure portion of the ULF wave is embedded between the old and new overshoots and forms an HSJ (Figure 2c). The length of this HSJ increases with time as the structure evolves downstream. As the HSJ interacts with the downstream magnetosheath plasma, the bent magnetic field lines within the HSJ become nearly perpendicular to its propagation direction. A bow wave with a mushroom-like morphology appears ahead of the HSJ during this compression process, as indicated by the red dashed curve in Figure 2d. Bow waves driven by fast magnetosheath jets and their associated electron energization have been reported in THEMIS and MMS observations \citep{Liu2019,Liu2020b}. Figures 2d-2e show another HSJ and bow wave formation sequence during the following shock reformation stage. The occurrence of bow waves at HSJ leading edges in the immediate downstream region appears to be related to both the injection phase of high dynamic pressure portions of upstream ULF waves and the stage of shock self-reformation. In Figure 2e, a new cycle from the old overshoot II to the newly forming overshoot III is shown. During this cycle, a new HSJ forms between approximately $X=595\ d_{i0}$ and $625\ d_{i0}$, together with a magnetosonic compression structure with a mushroom-like morphology at its leading edge. The formation of HSJs between old and new shock fronts during shock reformation is consistent with previous MMS observations \citep{Raptis2022}.

Figures 2f-2j show the corresponding evolution of the magnetic field component $B_y/B_0$. The distorted magnetic field lines in the downstream HSJ retain the curved magnetic topology of the upstream ULF wave. After the high dynamic pressure ULF wave crosses the shock, the transverse magnetic field component is amplified, and the amplitude of $B_y/B_0$ can exceed 6 in the immediate downstream region. Near the HSJ around $Y=70\ d_{i0}$, the HSJ highly squeezes the bent magnetic field into a thin current sheet on sub ion inertial scales, as highlighted by the red box in Figure 2i. Figures 2k-2o show the corresponding evolution of the current sheet at the foot of the bow wave. In Figure 2n, the local magnetic topology contains X-point and O-point structures in the $X$-$Y$ plane. In the following subsection, we focus on this localized HSJ leading edge and examine the reconnection signatures in the current sheet.

\subsection{Reconnection Signatures at an HSJ-driven Bow Wave} \label{sec:s3.3}

Figure 3 examines the current sheet formed near the leading edge of the HSJ-driven bow wave. Figures 3a-3c show the time evolution of the current density $J_z/J_0$ at $t=51$, 51.8, and $52.1\ \Omega_{ci0}^{-1}$, respectively. Magnetic field lines are shown as black curves with arrows. As the HSJ propagates from left to right, the oppositely directed magnetic field structures near the bow wave foot are compressed into a thin current sheet. The current sheet becomes progressively thinner from Figure 3a to Figure 3c, and its width is reduced to less than $1\ d_{i0}$ at $t=52.1\ \Omega_{ci0}^{-1}$.

To examine whether this current sheet shows reconnection signatures, we analyze the dissipation-related quantity $\mathbf{J}\cdot\mathbf{E}'$, where $\mathbf{E}'=\mathbf{E}+\mathbf{V}_e\times\mathbf{B}$ is the electric field in the electron frame \citep{Zenitani2011}. Figure 3d shows the total $\mathbf{J}\cdot\mathbf{E}'$ at $t=52.1\ \Omega_{ci0}^{-1}$. Positive values appear near the X-point, indicating local conversion from electromagnetic energy to plasma energy. These features are consistent with active energy conversion around the reconnecting current sheet center. The local current sheet coordinate system is shown in Figure 3c. Following the common LMN convention used in magnetic reconnection studies, $L$ is taken along the reconnecting magnetic field and the expected outflow direction, $N$ is the inflow or current sheet normal direction, and $M$ is the out-of-plane direction, approximately along the current direction \citep{Burch2016}. Figures 3e and 3f show the electron and ion bulk velocities along the $L$ direction, $V_{eL}/V_{A0}$ and $V_{iL}/V_{A0}$, after background subtraction. On both sides of the X-point, the electron and ion flows show bidirectional outflows along $\pm L$. The electron outflows are faster, whereas the ion outflows reach speeds comparable to $V_{A0}$ and occupy a broader region. These signatures suggest ion-coupled reconnection rather than electron-only reconnection. The combined signatures of an X-point topology, positive $\mathbf{J}\cdot\mathbf{E}'$, and bidirectional outflows of both species suggest that the compressed current sheet at the HSJ leading edge exhibits reconnection signatures. Such ion-coupled reconnection signatures associated with magnetosheath jet-driven bow waves are consistent with recent high-resolution MMS observations \citep{Ouyang2026}.

\subsection{Electron Energization in HSJ-Associated Reconnection} \label{sec:s3.4}

To examine electron energization in the HSJ-associated reconnecting current sheet, we first inspect the local electron energization rate $dU/dt$ at $t=52.1\ \Omega_{ci0}^{-1}$. As shown in Figure 4a, enhanced $dU/dt$ appears near the X-point of a reconnecting current sheet, indicating localized electron energization in this region. Figure 4a is obtained by evaluating the three terms on the right-hand side of Equation (1) locally at each grid point and taking their sum. To further check whether electrons are energized there, we sample electrons in a small box enclosing the X-point. Following previous full PIC studies of electron energization in magnetic reconnection \citep[e.g.,][]{Dahlin2014}, the energy of each sampled electron is calculated from its relativistic kinetic energy. For each electron, the Lorentz factor is defined as $\gamma_e=(1-v_e^2/c^2)^{-1/2}$, where $v_e^2=v_{ex}^2+v_{ey}^2+v_{ez}^2$, and the electron kinetic energy is given by $\varepsilon_e=(\gamma_e-1)m_ec^2$. Figure 4b shows the electron energy spectrum $f(\varepsilon_e)$ versus $\gamma_e-1$, i.e., $\varepsilon_e/\varepsilon_{e0}$, where $\varepsilon_{e0}=m_ec^2$. Here, $f(\varepsilon_e)$ is normalized by its maximum value, $f(\varepsilon_e)_{\max}$. The spectrum contains a thermal core and a high-energy nonthermal tail, highlighted in red, indicating that a fraction of electrons within the X-point region is energized above the core population. To identify the energization mechanisms, we employ the electron energization diagnostic based on guiding-center theory \citep{Northrop1963,Dahlin2014}.

\begin{equation}\label{eq1}
\frac{dU}{dt}
=
E_{\parallel}J_{\parallel}
+
\frac{p_{\perp}}{B}
\left(
\frac{\partial B}{\partial t}
+
\mathbf{u}_{E}\cdot\nabla B
\right)
+
\left(
p_{\parallel}
+
m_enu_{\parallel}^2
\right)
\mathbf{u}_E\cdot \boldsymbol{\kappa},
\end{equation}

where $U$ is the local electron kinetic energy density, so that $dU/dt$ represents the local electron energization rate, $E_{\parallel}$ is the parallel electric field, $J_{\parallel}$ is the parallel current density, $p_{\perp}$ and $p_{\parallel}$ are the perpendicular and parallel pressures, respectively, and $\mathbf{u}_E=\mathbf{E}\times\mathbf{B}/B^2$ is the $\mathbf{E}\times\mathbf{B}$ drift velocity. The curvature is $\boldsymbol{\kappa}=\mathbf{b}\cdot\nabla\mathbf{b}$, where $\mathbf{b}=\mathbf{B}/B$. This approach allows the electron energy gain to be separated into three terms on the right-hand side representing energization by the parallel electric field, perpendicular betatron energization, and Fermi energization, respectively. The adiabatic condition is satisfied over most of the analyzed region, except in very few localized areas, indicating that Equation (1) is broadly applicable (see the Appendix for details).

Based on this decomposition, Figures 4c-4e show the contributions from the three terms, respectively. In the X-point region, the betatron term appears to make a positive contribution and is comparable to $dU/dt$. This suggests that local electron energization near the X-point is associated with magnetic compression in the HSJ-associated reconnecting current sheet. Such betatron energization can occur when reconnected magnetic flux piles up in the outflow region, leading to local magnetic field amplification and an increase in the perpendicular adiabatic energy of electrons \citep{Fu2011,Huang2015,Wang2016}. Unlike in PIC simulations of Harris-type current sheets, electrons can be highly accelerated by the parallel electric field and Fermi-like mechanism \citep[e.g.,][]{Daughton2006,Fu2006,Huang2010}. \citet{Wetherton2022} demonstrated the balance between $E_{\parallel}$ and the other relevant physical terms. In guide-field reconnection, electron temperature anisotropy can become an important factor in balancing $E_{\parallel}$. Based on the decomposition of $dU/dt$ following \citet{Dahlin2014}, the parallel electric field acceleration, betatron heating, and Fermi acceleration all contribute to electron energization in the reconnection region in the present event.

 In comparison, only a weak signature of the parallel electric field term is visible near the X-point. The Fermi term shows limited contributions near the magnetic island edge and curved magnetic structures. Unlike in the island sea scenario \citep{Drake2013}, no frequent island coalescence or strong island contraction is evident in the present case. The limited Fermi acceleration may explain why the electron spectrum in Figure 4b develops a nonthermal tail rather than a power-law form.

\section{Conclusions and Discussion} \label{sec:s4}

Using 2-D full PIC simulations, we investigated how HSJs generated downstream of a parallel shock shape the topology of coupled downstream transients, including bow waves, compressed current sheets, and magnetic reconnection. The main results are summarized as follows.

1. Foreshock ULF waves are transmitted downstream across the shock, where their bent magnetic structures are further compressed at the leading edges of HSJs, forming thin current sheets and bow waves. These compressed current sheets exhibit reconnection signatures, including X-point topology, magnetic islands, enhanced $\mathbf{J}\cdot\mathbf{E}'$, and bidirectional Alfv\'enic outflows along the local reconnecting direction. The observed ion-scale outflows indicate that the compressed current sheets can develop into ion-coupled reconnection.

2. Electron energization occurs in the HSJ-associated reconnection region. The parallel electric field acceleration, betatron heating, and Fermi acceleration jointly contribute to the local electron energization, and the electron energy spectrum develops a nonthermal tail.

We also discuss how the ion-to-electron mass ratio affects HSJ formation in the full PIC simulations. Figure 5 shows the results for three mass ratios, $m_i/m_e=100$, 400, and 1200. Compared with the previous $m_i/m_e=1200$ simulation, no additional electron-scale instabilities were observed in the reconnection regions of interest in the $m_i/m_e=100$ and 400 simulations. The magnitudes and spatial distributions of the contributions from the parallel electric field, betatron acceleration, and Fermi acceleration remain almost unchanged (not shown here). These three processes jointly contribute to electron energization in the reconnection regions associated with HSJs. These results indicate that the lower-mass-ratio simulations adequately resolve the electron physics associated with HSJ-driven reconnection in this study. For $m_i/m_e=100$ (Figure 5a), the downstream HSJs readily fragment into small-scale structures, with both their lengths remaining relatively short (up to about $20\ d_{i0}$), similar to previous PIC simulations with $m_i/m_e=25$ \citep{Zhou2026}. When the mass ratio increases to $m_i/m_e=400$ (Figure 5b), the larger ion inertia allows the HSJs to become longer and more coherent. For $m_i/m_e=1200$ (Figure 5c), broader and more robust ion-scale HSJs are generated downstream, with lengths exceeding about $75\ d_{i0}$ within the simulated interval. These structures may extend further in longer simulations, but using longer simulation times or a more realistic mass ratio, such as $m_i/m_e=1836$, would substantially increase the computational cost and make the simulations infeasible. The alternating pattern between high-dynamic-pressure jets and low-dynamic-pressure cavities also becomes clearer. In three dimensions, such alternating jet-cavity topology may correspond to the honeycomb-like high- and low-dynamic-pressure structures reported in previous global hybrid simulations \citep{Ren2024a,Yang2024,Fatemi2024}. It is worth noting that the mushroom-like bow waves driven at the HSJ leading edges also become more visible at higher mass ratios, resembling the morphology found in previous hybrid simulations \citep{Karimabadi2014}. In summary, HSJs in the large mass ratio full PIC simulations are consistent with those in massless-electron hybrid simulations \citep[e.g.,][]{Karimabadi2014,Guo2022,Ren2024a}. We also performed calculations at a lower resolution and confirmed that the results were nearly identical (not shown).

An important limitation of the present study is the two-dimensional geometry of the simulations. Previous 3-D kinetic simulations have shown that reconnection may develop more complex flux-rope structures and instabilities than those represented by magnetic islands in 2-D. In particular, axial structure within flux ropes can enable energetic particles to escape from individual structures and return to the acceleration region, while flux-rope kink instability can produce strong field-line chaos and further enhance particle transport and acceleration \citep{Dahlin2017,Zhang2021}. In addition, astrophysical and space plasmas are generally turbulent, and 3-D turbulence can substantially alter the magnetic connectivity and reconnection dynamics relative to the 2-D case, including the reconnection rate \citep{Lazarian1999,Lazarian2020}. Therefore, some of the detailed reconnection properties obtained in the present 2-D simulations may not directly apply to three dimensions. Fully 3-D PIC simulations are beyond the computational scope of the present study and will be considered in future work.

More broadly, the termination shock is generally considered quasi-perpendicular near the heliospheric nose and becomes more oblique along the flanks, where it may act as a potential energetic particle source \citep{McComas2006,McComas2019}. Voyager observations show that ambient turbulence in the heliosheath is strong \citep{Zhao2025}, and such turbulence may locally modify the shock-normal angle and favor ``sea of islands" structures \citep{Zhao2020,Drake2013,Zank2015a,Zank2021}. Under outer-heliospheric conditions, kinetic simulations show that quasi-perpendicular shocks can also generate precursor ULF waves \citep{Lembege2020,Matsukiyo2024,Nakanotani2026}, which may play an important role in downstream particle acceleration, which remains a topic for future study.

\section*{Acknowledgments}
We are grateful to S. Zenitani at Space Research Institute, Austrian Academy of Sciences, M. Swisdak of the University at Maryland and J. Giacalone at the University of Arizona for their valuable suggestions. This work is jointly supported by the National Key R\&D Program of China (2025YFF0512102, 2021YFA0718600), the National Natural Science Foundation of China (42150105, 42188101 and 42274210), the Climbing Program of NSSC (E6PD3012), and the Specialized Research Fund for State Key Laboratories of Solar Activity and Space Weather. H.H. is supported by the Royal Society award \verb|URF\R1\180671|. S.M. is supported by Grant-in-Aid for Scientific Research (B)23K22558 from JSPS. C.H. appreciates the financial support by the Key Research Program of the Chinese Academy of Sciences (KGFZD-145-2023-15).

\clearpage
\begin{figure}[ht!]
\centering
\includegraphics[width=\textwidth,height=0.72\textheight,keepaspectratio]{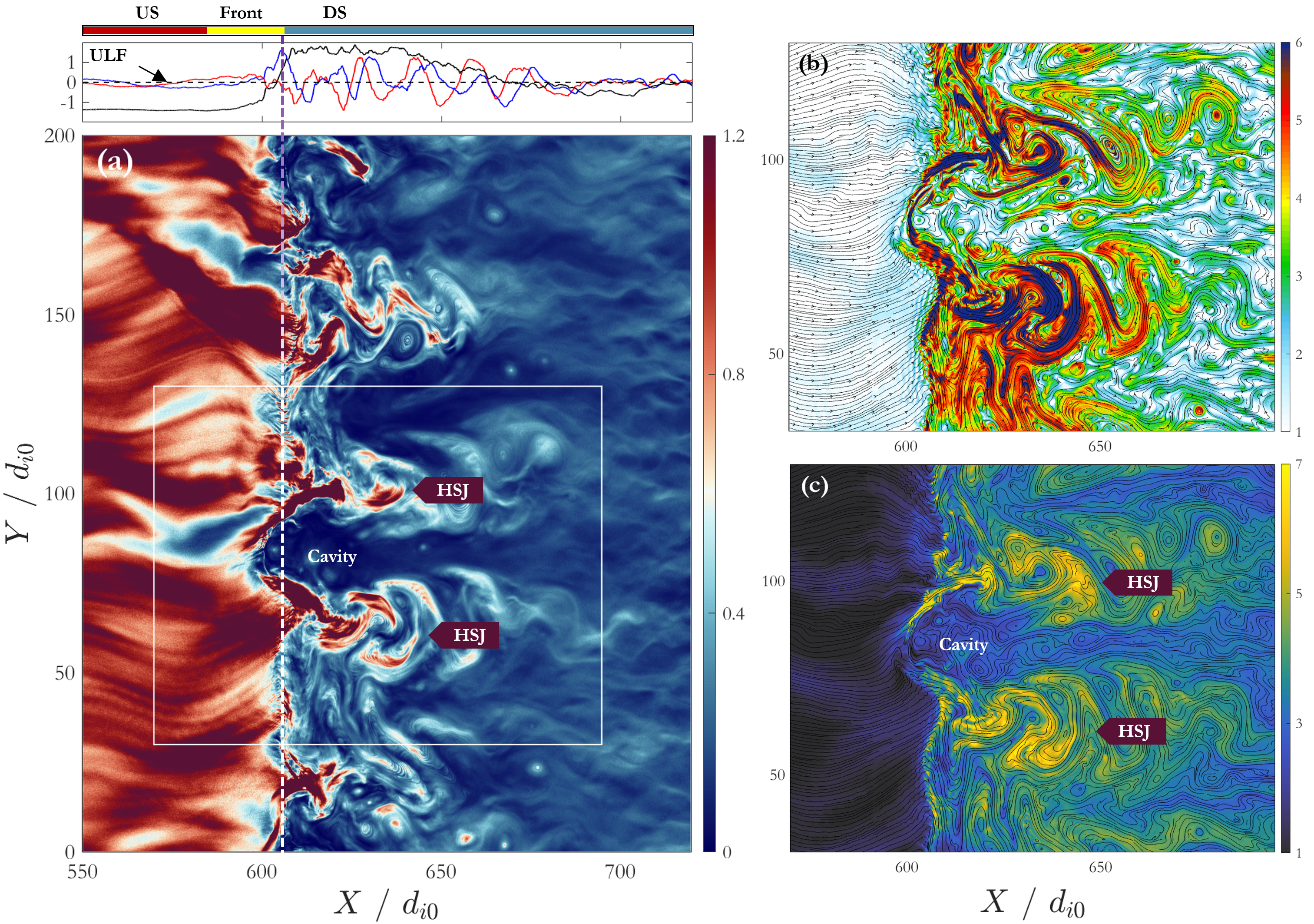}
\caption{\label{fig:fig1}{Overview of HSJs at a parallel shock. (a) Ion dynamic pressure $P_d/P_{d0}$ at $t=54\ \Omega_{ci0}^{-1}$. Warm colors highlight HSJs, and cool colors mark cavities. The y-averaged $B_y$ (red), $B_z$ (blue), and shifted B (black) are shown above panel (a). Horizontal bars mark the upstream (US), shock front (Front), and downstream (DS) regions from left to right. The vertical dashed line indicates the location of the shock front. Panels (b) and (c) show enlarged views of the white box in panel (a), with $B/B_0$ and $N/N_0$, respectively. Magnetic field lines are shown as black curves for reference.}}
\end{figure}
\FloatBarrier

\clearpage
\begin{figure}[ht!]
\centering
\includegraphics[width=\textwidth,height=0.72\textheight,keepaspectratio]{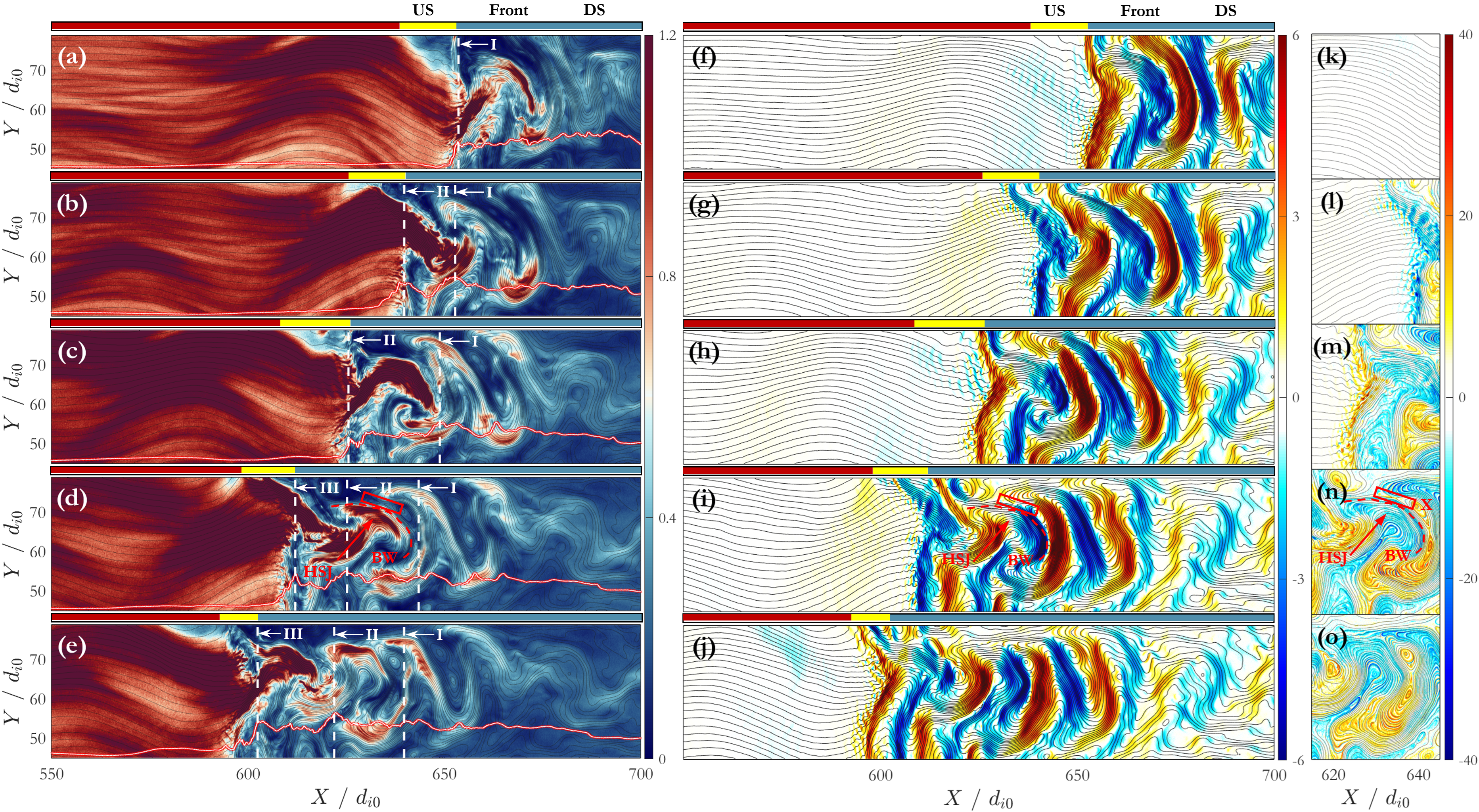}
\caption{\label{fig:fig2}{ULF-wave compression and HSJ formation at a parallel shock. Panels (a)-(e) show the time evolution of the ion dynamic pressure $P_d/P_{d0}$ at $t=40.2$, 44.2, 48.2, 52.9, and $56\ \Omega_{ci0}^{-1}$, respectively. In each panel, the red, yellow, and blue horizontal bars at the top mark the upstream (US), shock front (Front), and downstream (DS) regions at the corresponding time. The white curve with a red outline shows the ion number density $N_i/N_0$ averaged over $Y=65$--$75\ d_{i0}$. The white vertical dashed line denotes the overshoot position of the shock front. The shock front can undergo self-reformation; labels I, II, and III indicate the old and new shock overshoots during the self-reformation cycles. Panels (f)-(j) show the corresponding profiles of the magnetic field component $B_y/B_0$. Panels (k)-(o) represent the current sheets at the leading edge of the HSJs downstream of the shock. Magnetic field lines are also shown as black curves for reference.}}
\end{figure}
\FloatBarrier

\clearpage
\begin{figure}[ht!]
\centering
\includegraphics[width=\textwidth,height=0.72\textheight,keepaspectratio]{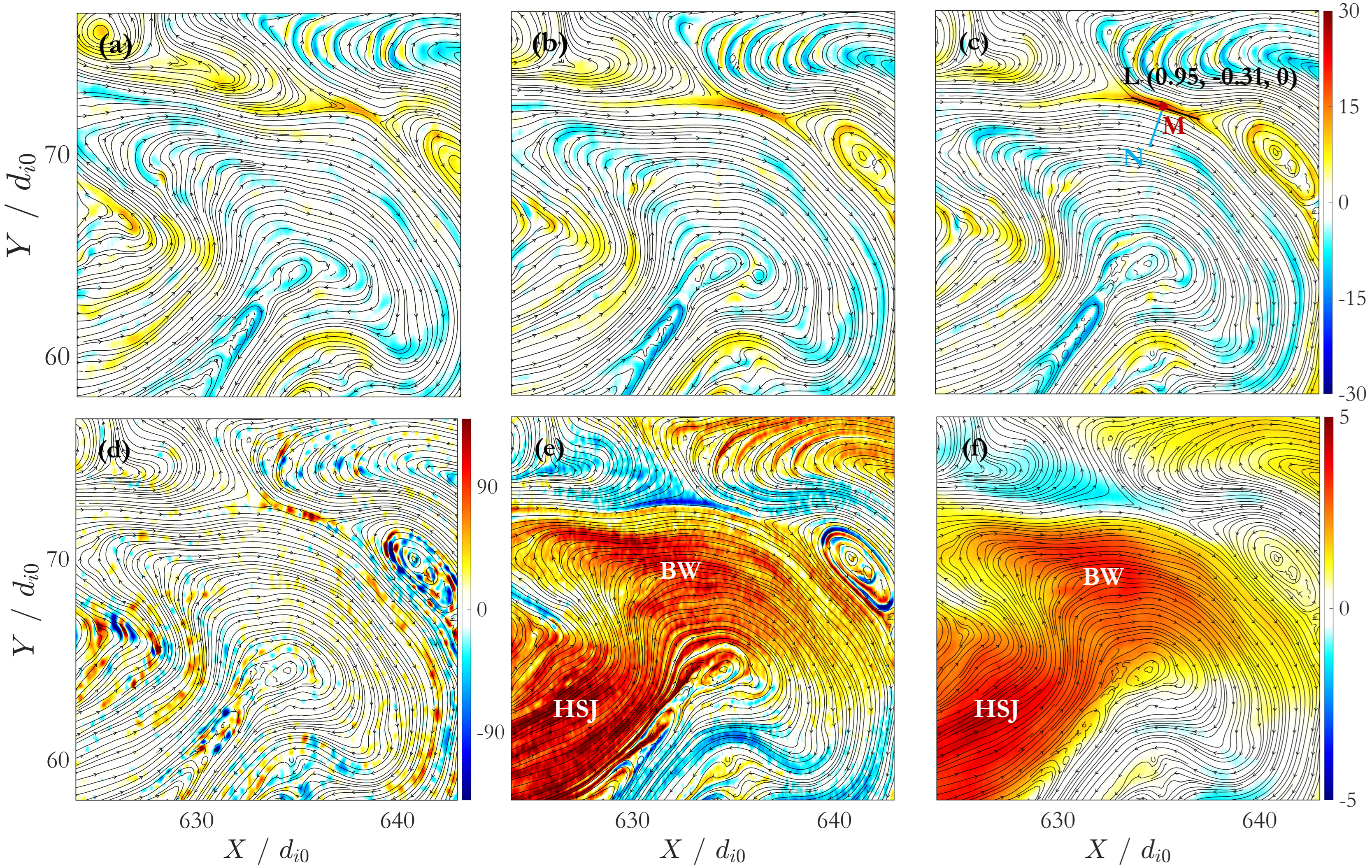}
\caption{\label{fig:fig3}{Current sheets with reconnection signatures at HSJ-driven BWs. Panels (a)-(c) show the evolution of the out-of-plane current density $J_z$ at $t=51$, 51.8, and $52.1\ \Omega_{ci0}^{-1}$, respectively. The L-M-N directions in panel (c) mark the local current sheet coordinate system near the X-point, with N being the normal direction. Panel (d) shows the total dissipation-related quantity $\mathbf{J}\cdot\mathbf{E}'$ at $t=52.1\ \Omega_{ci0}^{-1}$. Panels (e) and (f) show the electron and ion bulk velocities (background subtracted), $V_{eL}/V_{A0}$ and $V_{iL}/V_{A0}$, respectively, at the same time. Magnetic field lines are shown as black curves for reference.}}
\end{figure}
\FloatBarrier

\clearpage
\begin{figure}[ht!]
\centering
\includegraphics[width=\textwidth,height=0.72\textheight,keepaspectratio]{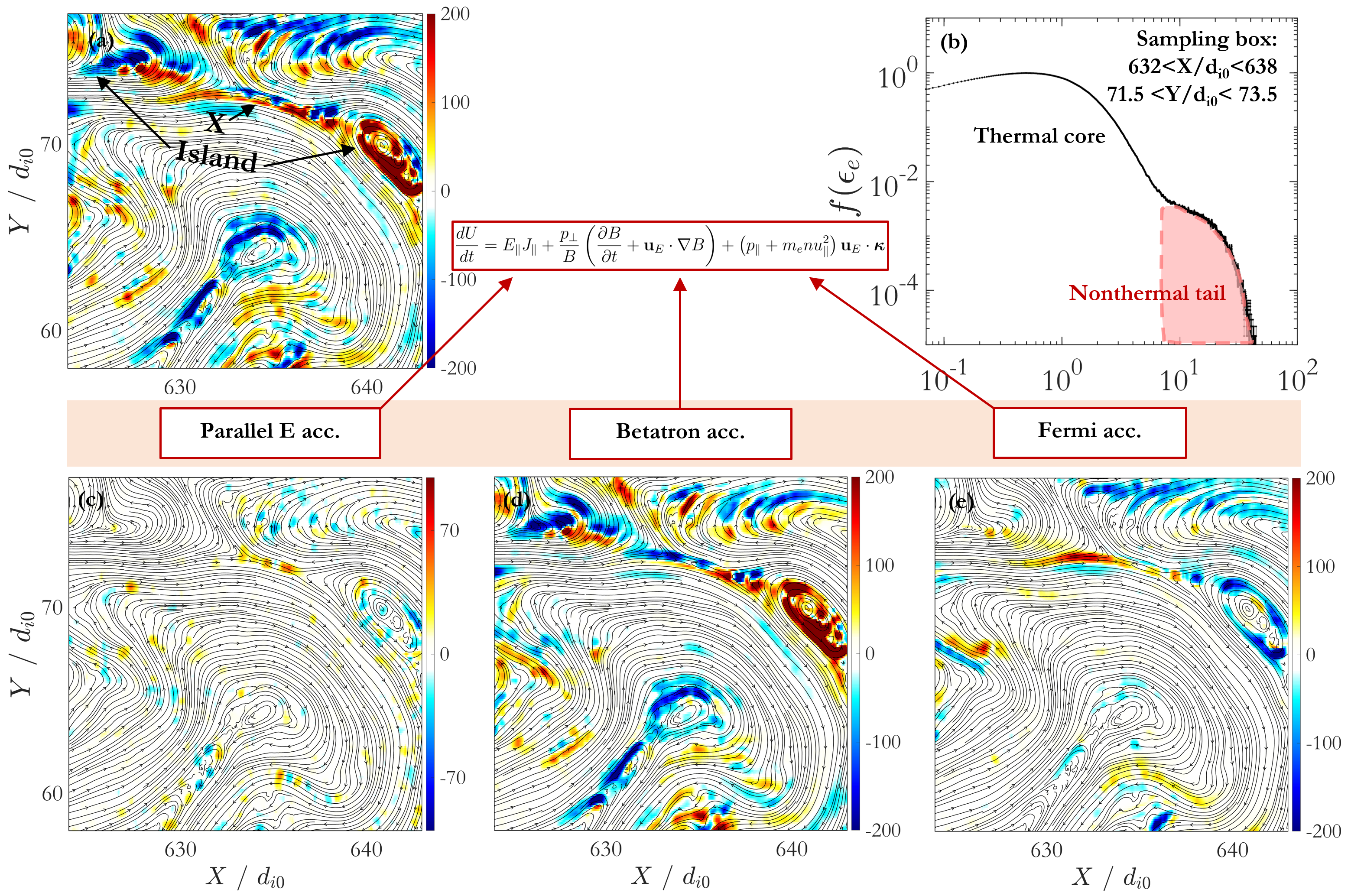}
\caption{\label{fig:fig4}{Electron energization in HSJ-associated reconnecting current sheets. All panels correspond to $t=52.1\ \Omega_{ci0}^{-1}$. (a) The local electron energization rate $dU/dt$. (b) The electron energy spectrum sampled in a box around the X-point, with the nonthermal tail highlighted in red. (c)-(e) The contributions from the parallel electric field term, the betatron term, and the Fermi term, respectively. Magnetic field lines are shown as black curves for reference.}}
\end{figure}
\FloatBarrier

\clearpage
\begin{figure}[ht!]
\centering
\includegraphics[width=\textwidth,height=0.72\textheight,keepaspectratio]{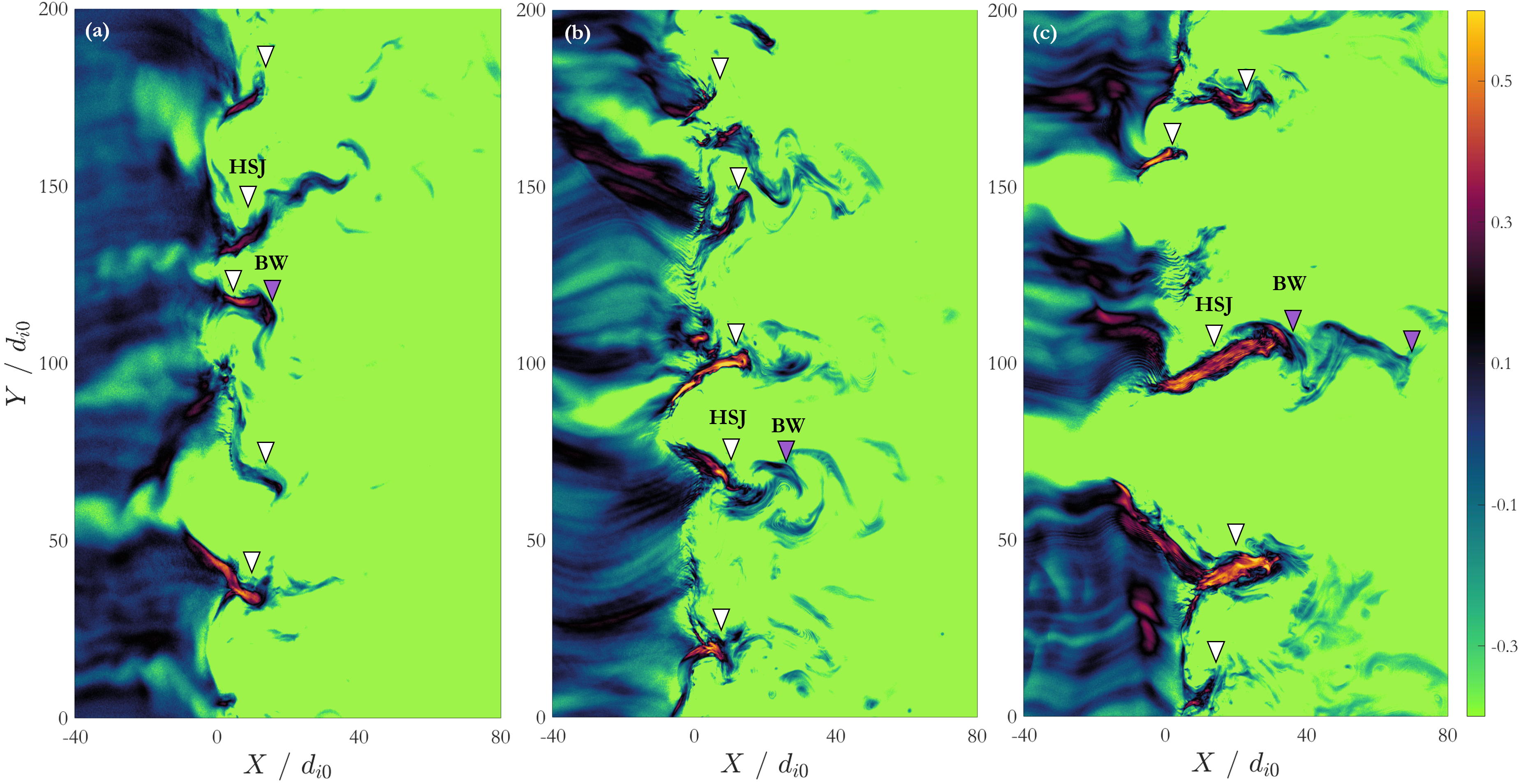}
\caption{\label{fig:fig5}{Effect of the ion-to-electron mass ratio on simulated HSJs. Panels (a)-(c) show simulations with mass ratios $m_i/m_e=100$, 400, and 1200, respectively. The white and purple arrowheads mark HSJs and BWs, respectively. The color scale shows $\log_{10}(P_d/P_{d0})$. The shock fronts are shifted to $X=0$ for comparison.}}
\end{figure}
\FloatBarrier

\clearpage
\appendix
\section{Electron Agyrotropy}

Figure A1 shows the electron agyrotropy, $A\Phi_{e}$, calculated for the reconnection region shown in Figure 4, following the method of \citet{Scudder2008}. Here, $A_{\Phi e}$ is defined as
$$
A\Phi_{e}=2\frac{|P_{\perp e1}-P_{\perp e2}|}{P_{\perp e1}+P_{\perp e2}},
$$
where $P_{\perp e1}$ and $P_{\perp e2}$ are the two pressure eigenvalues associated with eigenvectors perpendicular to the mean magnetic field direction. The results show that $A\Phi_{e}$ is below 0.2 over most of the region, indicating weak electron agyrotropy and approximately adiabatic electron behavior. This supports the broad applicability of Equation (1) and the present interpretation of electron energization over most of the region \citep{Dahlin2014}. Very localized nonadiabatic effects may nevertheless occur near the X-point in the turbulent downstream, and the present interpretation may evolve as these effects are examined in future analyses.

\renewcommand{\thefigure}{A\arabic{figure}}
\renewcommand{\theHfigure}{A\arabic{figure}}
\setcounter{figure}{0}
\begin{figure}[ht!]
\centering
\includegraphics[width=0.75\textwidth,height=0.72\textheight,keepaspectratio]{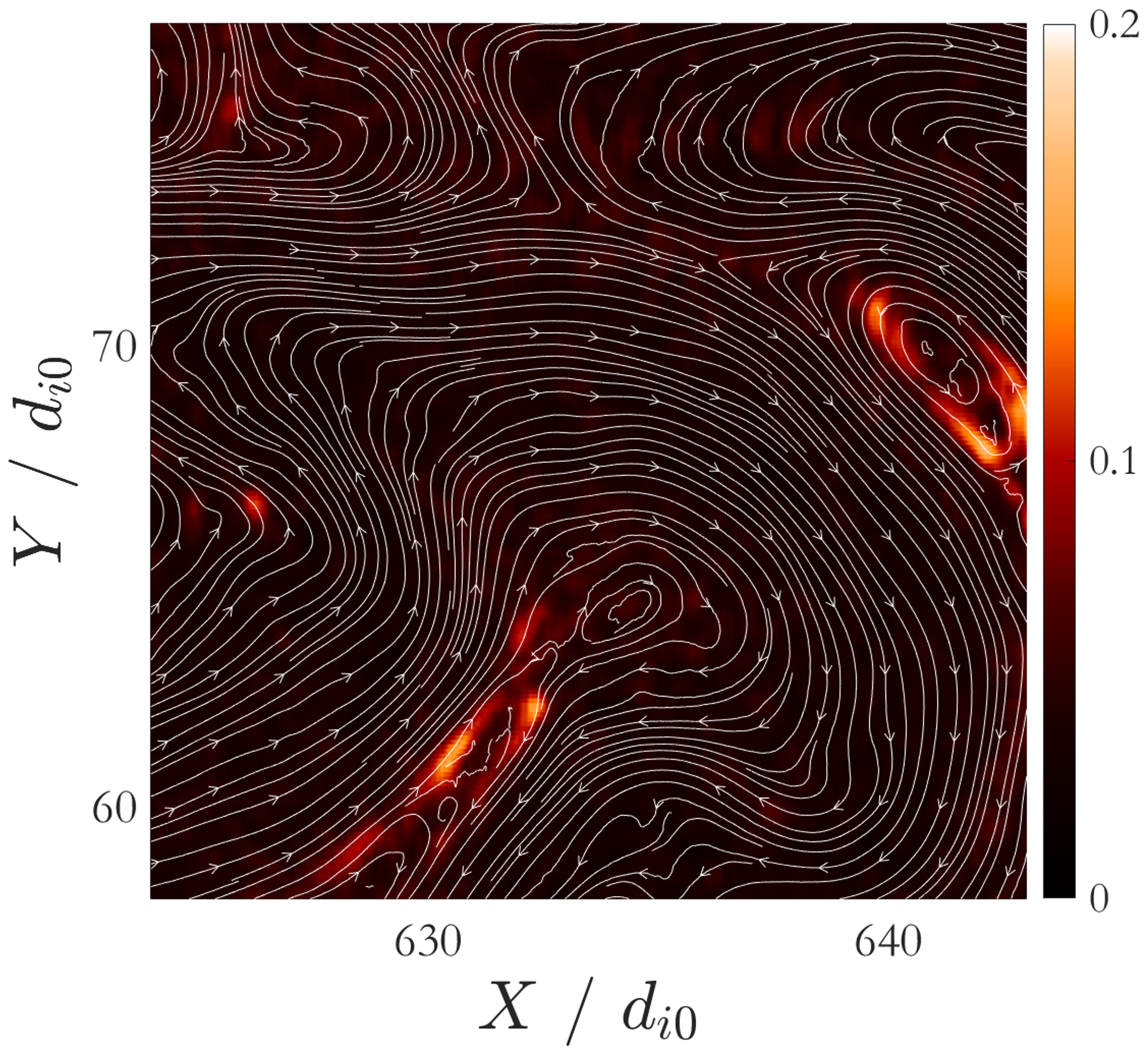}
\caption{\label{fig:figA1}{Agyrotropy calculated for the reconnection region shown in Figure 4, following the method of \citet{Scudder2008}.}}
\end{figure}
\FloatBarrier

\clearpage
\bibliographystyle{unsrtnat}
\bibliography{references}

\end{document}